\documentstyle[aps,prb,epsf,twocolumn]{revtex}
\begin{document}
\draft
\title{ Magnetization and Level Statistics at Quantum Hall Liquid-Insulator 
Transition
in the Lattice Model}
\author{M.~Ni\c{t}\~a$^{1}$ and A.~Aldea$^{2}$}
\address{$^1$National Institute of  Materials Physics, POBox MG7,
Bucharest-Magurele, Romania}
\address{$^2$Institut f\"{u}r Theoretische Physik, Universit\"{a}t zu K\"{o}ln,
D-50937 K\"{o}ln,Germany}
\maketitle
\begin{abstract}
Statistics of level spacing and magnetization are studied for the
 phase diagram of
the integer quantum Hall effect in a 2D finite lattice model with
 Anderson disorder. 
\end{abstract}

The way in which the increasing disorder induces the insulating
state when starting from the integer quantum Hall (IQH) state is a topic of
controversy
between the continuum and lattice models of the 2D electronic gas
in strong magnetic field. The continuum approach predicts 
the crossover between the adjacent quantum Hall plateaus, ending up with 
the insulating state when the degree of disorder increases 
or, equivalently, the limit of small magnetic field is  considered.
This is
due to the so-called 'floating up' of the critical energies $E_{c}$
which occurs with increasing disorder.  
($E_{c}$ is the energy where
the localization-delocalization transition takes place
in the thermodynamic limit).
In the critical region, i.e., when the Fermi
energy $E_{f}$ crosses an extended state energy  $E_c$, 
the transverse (Hall) conductivity  is $\sigma_{xy}^c=\nu-1/2$
(at the transition  between the plateaus $\nu$ and $\nu-1$)\cite{1}. 
This means that at large disorder (or low field), the cascade of 
transitions must end
with $\nu=1\rightarrow\nu= 0$ (insulator). The experiments give controversial 
information in what concerns the possibility to observe this
last transition (see Ref.2 and the references therein, Ref.3)
The sensible conclusion
can be found in Ref.2 and Ref.4 suggesting that the theoretical 
results of the 
scaling theory, which are obtained for zero temperature and infinite systems,
cannot be checked easily by experiments which are done for finite samples
and at low (but 
nevertheless finite) temperature.
The  evaluation of the critical value of the longitudinal 
conductivity is also a difficult task.
Lee, Kivelson and Zhang show in the frame of {\it corresponding states law} 
that $\sigma_{xx}^c=1/2$ for any $\nu$; approaching the question in 
the opposite way, Zirnbauer assumes $\sigma_{xx}^c=1/2$ and finds
agreement with the numerical simulations. The numerical calculation performed 
by Huo, Hetzel and Bhatt for the lowest Landau level produces 
also $0.5$ ; Huckestein and Backhaus  show that this value is  correct
even in the presence of the electron-electron interaction \cite{5}.

%-----------------------------------------------------------------

More recently, the same problem has  been approached also in  lattice
models. The results are again controversial, since Yang and Bhatt
\cite{6}, in a discussion based on the calculation of the Chern numbers
find a tiny floating-up of the extended states, while Xie at al \cite{7}
do not find any changing in the position of extended states and
affirm that the floating picture is not valid in this model.
%do not find anything like that. 
The type of boundary conditions used in the 
lattice model is important; the above mentioned authors use periodic
conditions. The Dirichlet conditions around the plaquette, 
which represent a more physical situation,
%seams to be more 
%physical and 
are used by Sheng and Weng \cite{8}; their approach, consisting
in numerical calculation of the  conductance
performed  over ensembles of disordered finite plaquettes, indicates that
%(described by the Hamiltonian Eq.(1)) 
the QH-insulator transition may occur from any IQH plateau of 
index $\nu$ directly to 
the insulator. In the middle of the transition region the
conductances satisfy the relation 
$<\sigma_{xx}>=<\sigma_{xy}>=\nu/2$ (here $<...>$ means ensemble average)
and the localization length diverges; this region is called 'metallic'.

%------------------------------

In this context we study some new relevant features of the
disordered lattice model in magnetic field with vanishing boundary
conditions, the attention being paid 
especially to magnetization, level spacing distribution and critical 
conductances in the metallic region.
%(The interesting aspects arise from the  presence
%of  two types of electronic states -bulk and edge- 
%and the modification of their 
%%spectral, magnetic and transport
%properties due to disorder.)
%and their modifications due to the disorder.

The discussion is based on the spin-less one-electron
Hamiltonian in  perpendicular magnetic field 
defined on a 2D square lattice
with N sites in one direction and M sites along the other one, which reads  
as follows:
%\begin{equation}
$$H=\sum_{n=1}^{N}\sum_{m=1}^{M} [~\large{\epsilon}_{nm} \vert n,m
\rangle\langle n,m\vert + 
^{i2\pi\phi m}\vert n,m\rangle\langle n+1,m\vert $$ 
$$+\vert n,m\rangle\langle n,m+1\vert + h.c.~]~~~~(1) $$ \\
%\end{equation}
where ${\vert n,m \rangle}$ is a set of orthonormal  states, localized at the 
sites (n,m), and $\phi$  is the magnetic flux through the unit cell measured
in quantum flux units. 
In Eq.(1) the hopping integral at $\phi=0$ is
taken unity serving as the energy unit and the diagonal energy
$\epsilon_{nm}$ is a random variable  distributed according to the
probability density:
$$ P(\epsilon)= \cases {1/W~, ~~~~    -W/2 <\epsilon < W/2 \cr
    0~ ,~~~~~~~~~~~~~ otherwise,~~~~~~~~~(2)}$$                      
(having zero mean $\bar\epsilon=0$).

%----------------------------------------------------

The averaged spectrum of Hamiltonian (1) is depicted in Fig.1  
for $\phi = 1/10$ and the disorder amplitude in the range  $W\in[0,10]$.
%the disorder amplitude being in the range  $W\in[0,10]$ . 
The lines represent the  mean eigenvalues $<E_n>$ as function of disorder
amplitude $W$.

In order to study the phase diagram we calculate the longitudinal and Hall
conductances of this system 
at constant magnetic flux and varying disorder.
The different phases: quantum Hall, metallic, insulating
are  characterized not only by conductance but also by the 
specific distribution of the level spacing
and by the  current density on the plaquette, 
described by the operator:
$$~~~~{\bf J}_{nm}^{n'm'}=i~t_{nm}^{n'm'}~({\bf r}_{nm}-{\bf r}_{n'm'})~
\vert nm\rangle\langle n'm'\vert~+h.c   ~~~~~~(4)$$
%which will be calculated on different eigenvectors of the Hamiltonian (1)
(here~$t_{nm}^{n'm'}$ is the hopping integral between the sites
${\bf r}_{nm}$ and ${\bf r}_{n'm'}$).

%-------spectrum+magnetization----------------------------------
%f1
\begin{figure}
\vspace{-2.4cm}
\epsfxsize 50mm
\epsffile{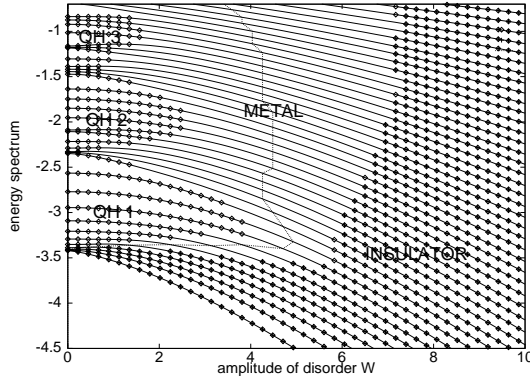}
\caption{Phase diagram of IQH effect. N=M=10
. The insulator
regime is depicted for  $\sigma_{xx} < 0.2$ (in $e^2/h$ units) and the metal
 regime
for $\sigma_{xx} > 0.2$ . In the QH regime  
$\sigma_{xy} =$ integer and  $\sigma_{xx}$ is negligible. The dotted line
in the metallic region
corresponds to the critical points ($\sigma_{xx} = \sigma_{xy}$). }
\label{Fig.1}

\end{figure}

It is opportune to  remind previously that for a {\it clean} system
($\epsilon_{nm}=0$) with cyclic boundary 
conditions (i.e for a torus) and commensurate values of the magnetic
flux through the unit cell, the spectrum consists of degenerate bands
separated by gaps (the well-known Hofstadter butterfly). 
However, when vanishing boundary conditions are imposed (i.e., for plaquette 
\cite{A} or cylinder\cite{Sch} geometry) the gaps 
get filled with 'edge states', localized  close to the edges of the 
sample. The other states, the 'bulk' ones, remain grouped in
bands on the energy scale, while geometrically are concentrated in the 
middle of the plaquette.
The two types of states differ also by their chirality,
i.e. by the sign of the derivative $ dE_{n}/d\phi$.
%f2
\vspace{-3.2cm}
\begin{figure}
\epsfxsize 80mm
\epsffile{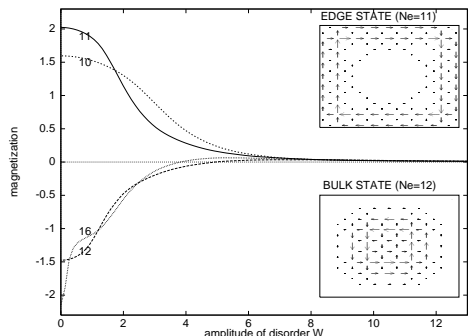}
\vspace{-3cm}
\caption{The decay of the magnetization (in arbitrary units) 
with increasing disorder for the
edge states No. 11 and 10 and for the bulk states No. 16 and 12.
The distribution of current at $W = 0$ is shown in the insets.
$\phi=1/10$.}
\label{Fig.2}
\end{figure}

The effect on the orbital magnetization of each state is immediate:
the expectation values of the operator
${\bf M}=\int({\bf r} \times {\bf j}({\bf r}))~ dS$  calculated
on the eigenstates of the Hamiltonian (1) have different signs depending on
whether the state is bulk- or edge-type. 

Fig.2 shows that the magnetization 
of the edge eigenstate No.11 is positive $M_{11}>0$, but 
the bulk eigenstate No.12 has $M_{12}<0$ ; the local currents
corresponding to the two states are also shown in insets.
In the same figure one anticipates that the increasing disorder produces a 
monotonic decrease of the magnetization.
The magnetization of all states in the spectrum
is shown in  Fig.3a and b, 
for the clean and disordered system, respectively
(the well-known electron-hole symmetry of the Hofstadter spectrum
is evident also in the aspect of the magnetization).
The disorder effect consisting in the broadening of the bands 
and narrowing of the gaps is obvious in the second figure.
More notable is the magnetization of the ground state $M_g$
which can be compared successfully with  experimental results.
Assuming that the spectrum is filled up to the Fermi energy $E_f$,
due to the alternating sign of $M_n$ in different regions of 
the spectrum, the quantity
$$M_{g}=\sum_{(E_{n}<E_{f})} M_n~~~~~~~~~~~~~~~~~~~~~~~~~~~~~~~~~~~~~(5)$$
shows, as function of the number of occupied states, a sawtooth aspect
(see Fig.4) which is the same as in the experiments by Wiegers
at al \cite{11}, including the fact that the jumps of $M_{g}$  occur
at the center of the gaps.  In our model, the number of teeth depends
on the number of gaps that can be resolved. 
For modulated quantum wires such a sawtooth aspect was obtained also 
for the thermodynamic magnetization \cite{12}.
%f3

\begin{figure}
\vspace{-5.0cm}
\epsfxsize 65mm
\epsffile{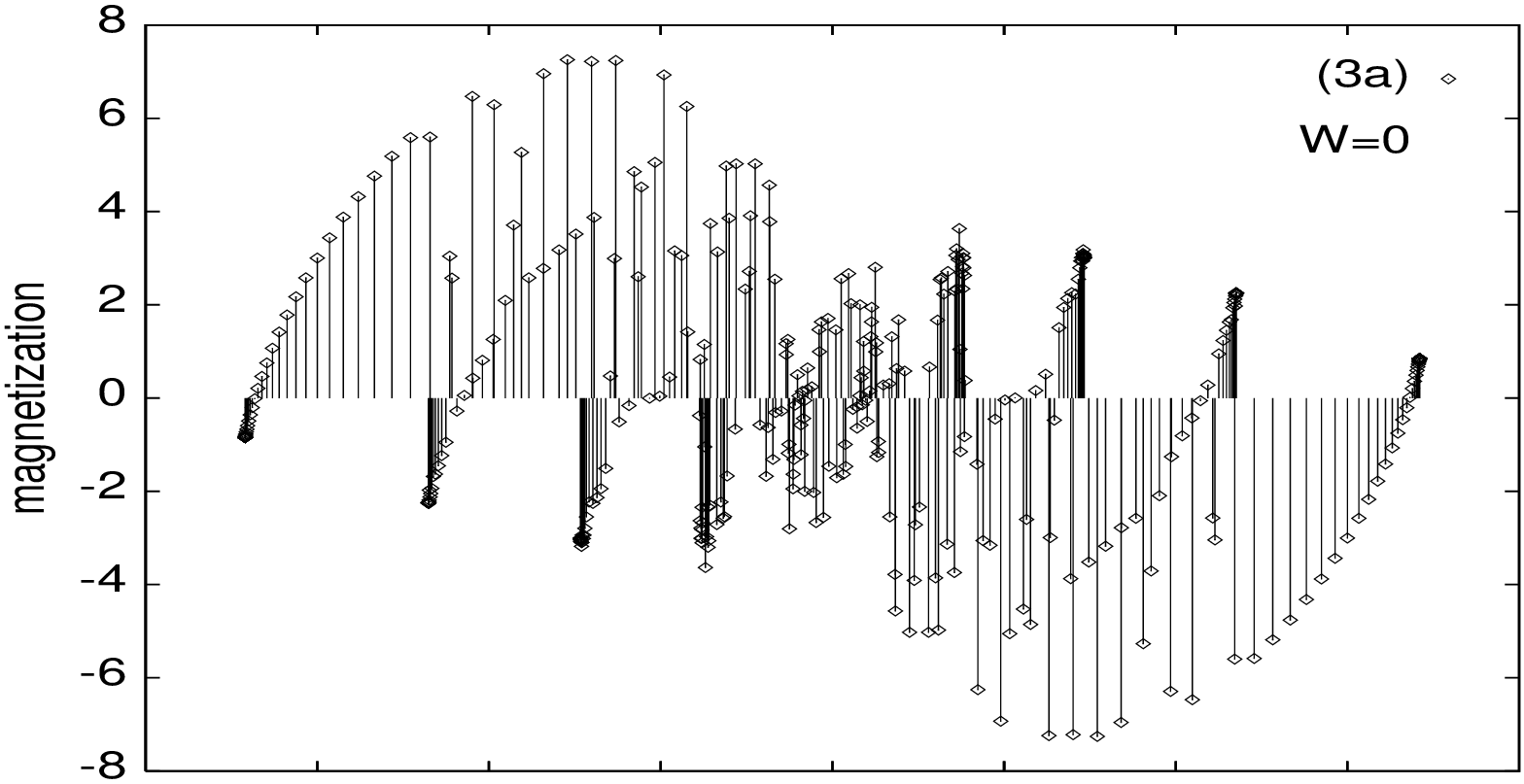}
\vspace{-4.7cm}
\epsfxsize 65mm
\epsffile{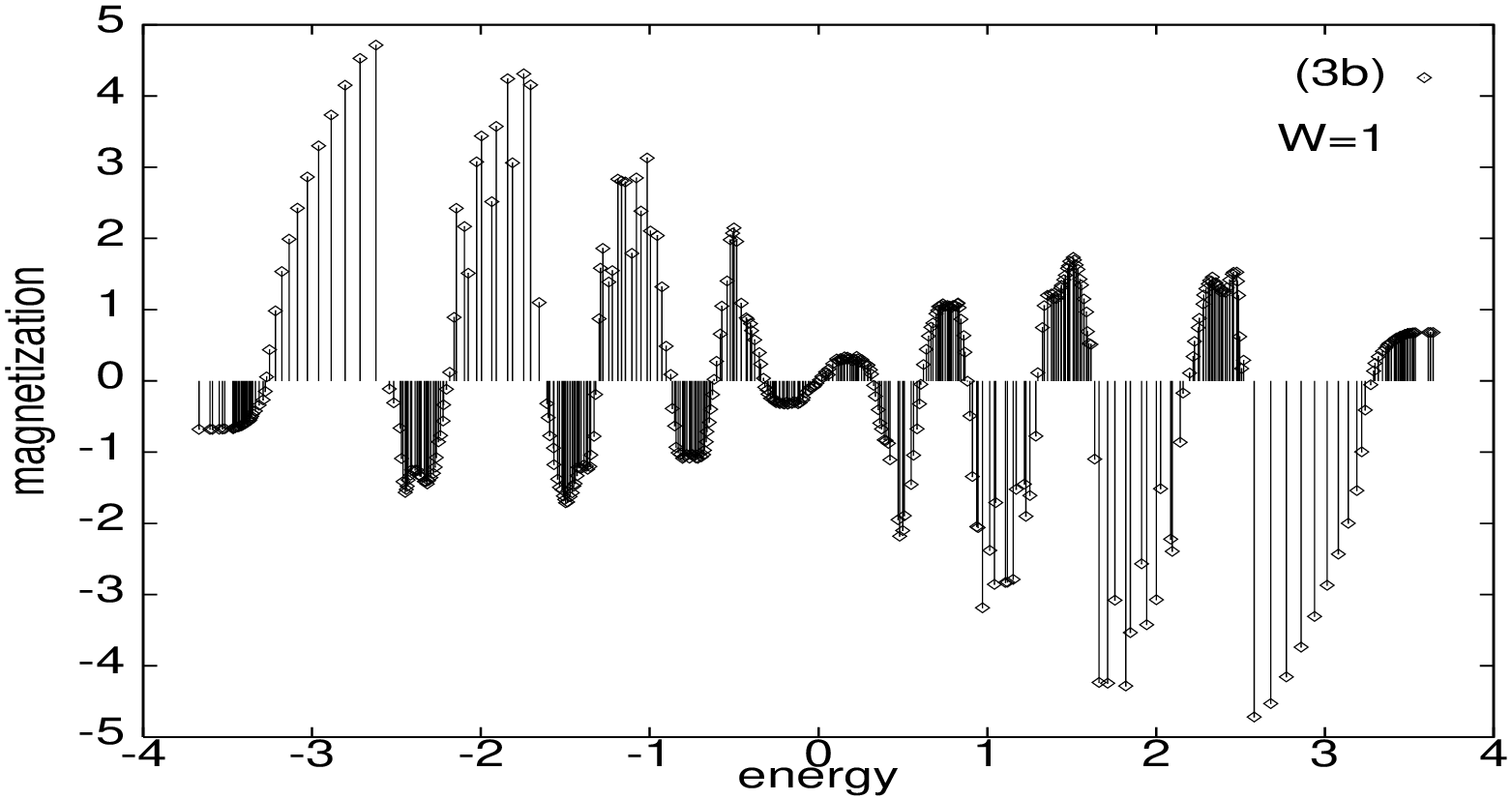}
\caption{Magnetization (in arbitrary units) vs. the energy. $W = 0$ in
fig.3a and 
$W = 1$ in fig.3b. 
N=M=20. $\phi=1/10$.}
\label{Fig.3}

\end{figure}
%f4
\vspace*{-4.7cm}
\begin{figure}
\hspace{-3.1cm}
\epsfxsize 115mm
\epsffile{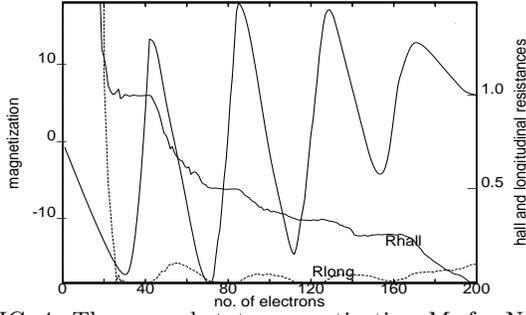}
\vspace{-7.0cm}
\caption{The ground state magnetization $M_g$  for N=M=20 and $W = 1$
vs. the number of electrons
$N_e$ (left scale). $R_{xx}$ and $ R_{xy}$
for the same plaquette coupled to leads (right scale).}
\label{Fig.4}

\end{figure}

%-----------------fluctuations----------------------------------
%f5

\vspace{0.5cm}
\begin{figure}
\epsfxsize 75mm
\vspace{-3.4cm}
\epsffile{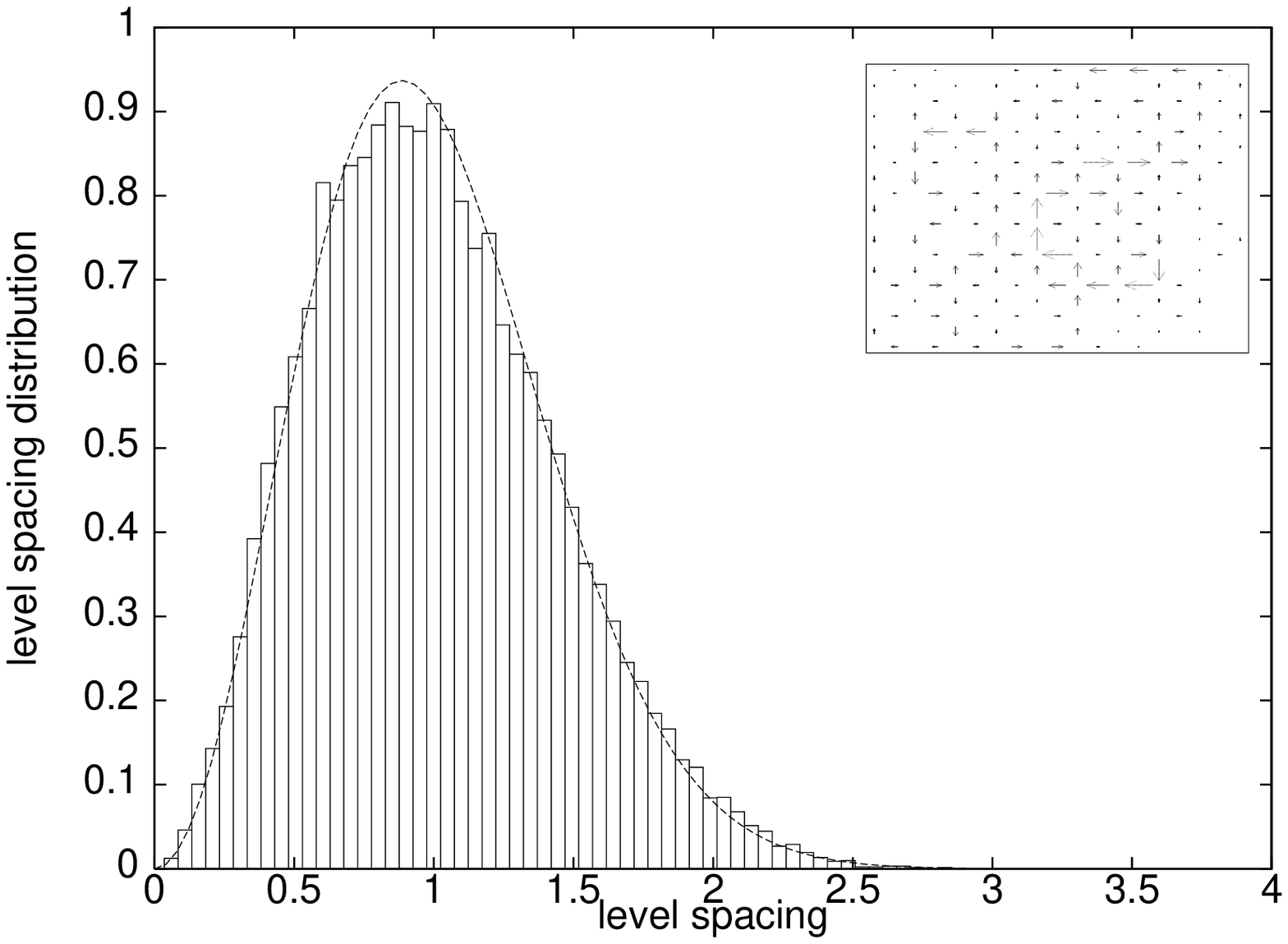}
\epsfxsize 75mm
\vspace{-5.5cm}
\epsffile{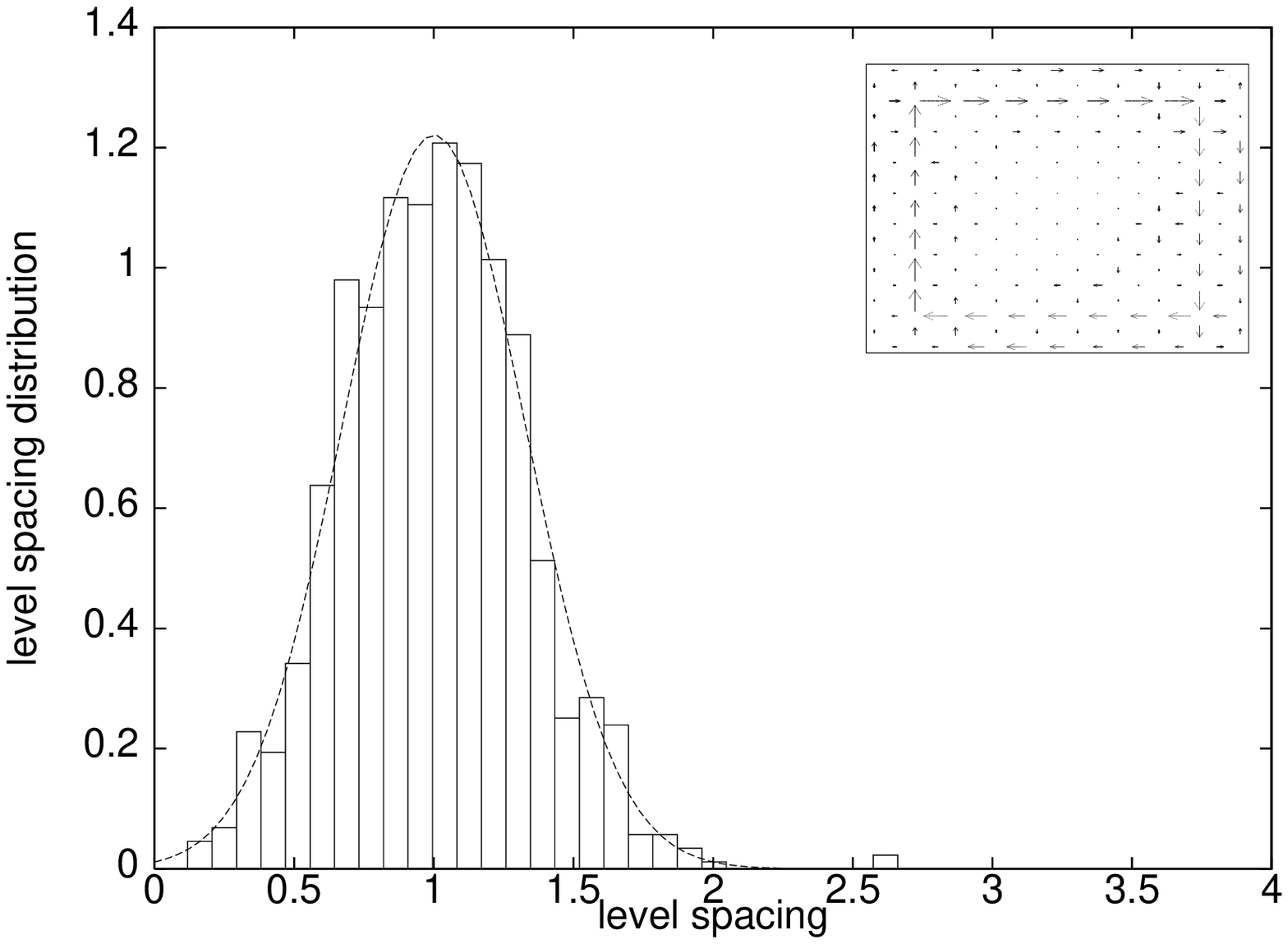}
\epsfxsize 75mm
\vspace{-5.5cm}
\epsffile{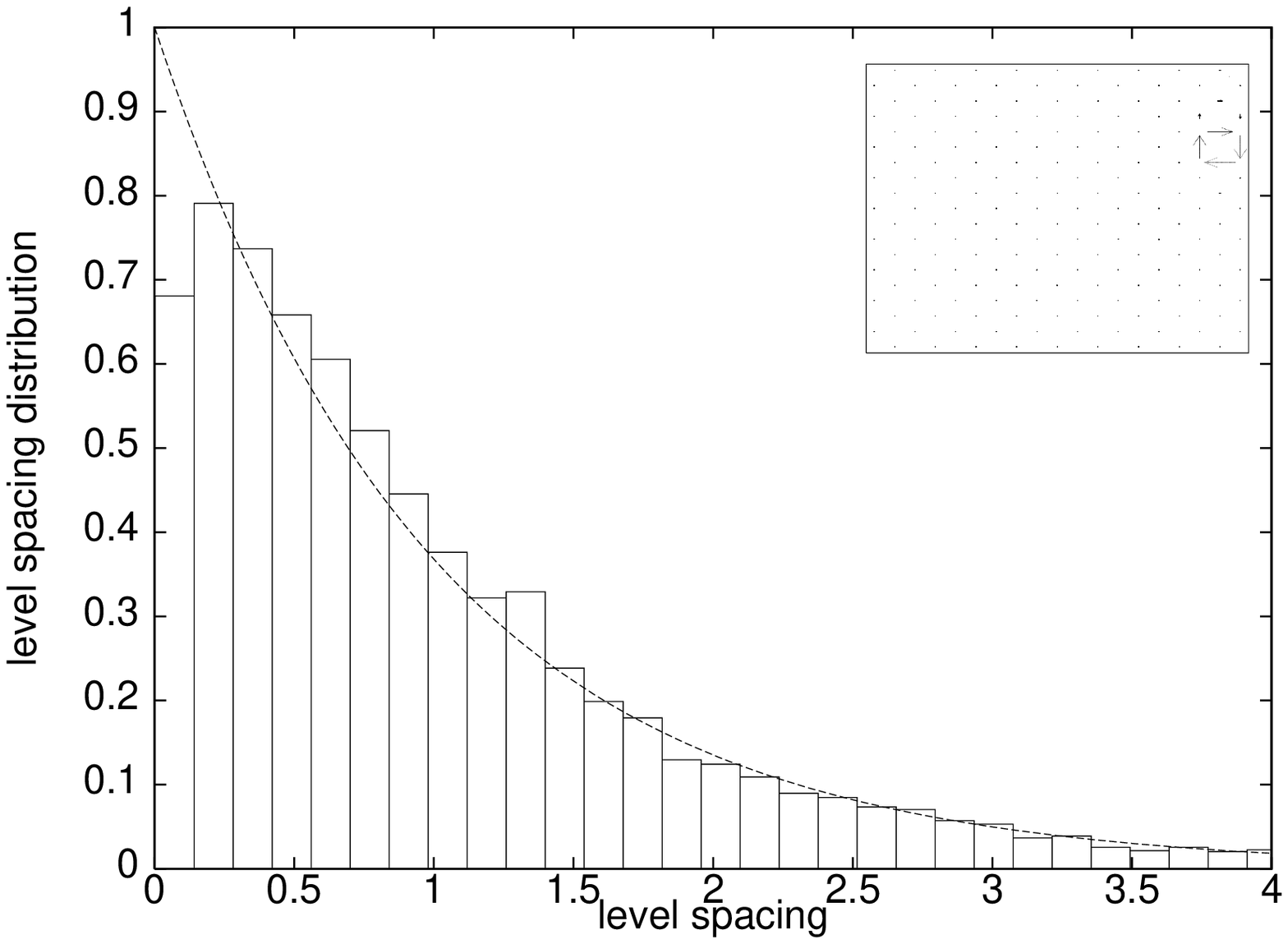}
\vspace{-2.9cm}
\caption{Level spacing distribution $P(t)$ for three typical situations.
Fig.5a gives the  distribution of level spacing for $E \in [-2.4,-1.3]$
and $W=3$. The dotted line is the 
Wigner-Dyson distribution
with $\beta=2$.
In fig.5b  
$E \in [-2.9,-2.7]$ and $W=1$. The dotted
line is the Gaussian function whose variance equals the  calculated variance
of the histogram $\delta(t)=0.33$.
In fig.5c $E \in [-8,-6]$, $W=13$ and $\delta(t)=0.92$. 
The dotted line is the Poisson function.
The corresponding typical current distributions are shown in the insets.
N=M=10 and the number of disorder configurations is equal to 1000.
 $\phi=1/10$.}
\label{Fig.5}
\end{figure}

When the  Anderson potential (Eq.2) is switched on and $W$ is increased
continuously, the bands become broader and broader, the disorder
spreads  the states over the whole plaquette, giving rise to extended
disordered states, and finally produces a quasi-continuum of 
localized states; even the edge states, which are more robust, disappear 
gradually into the quasi-continuum.
The nature of the states can be checked by calculating the distribution
of level spacing for various degrees of disorder, in different domains
of the spectrum. 
Let  $s_{n}$ be the level spacing between two consecutive eigenvalues
$E_{n}$ and $E_{n+1}$ and  define $t_{n}= s_{n}/\langle s_{n}\rangle$,
where $\langle s_{n}\rangle$  is the mean level spacing.

In Fig.5 we have three typical distribution functions $\it P(t)$
relevant for different values of disorder and energy interval.
The Wigner-Dyson (WD) surmise with $\beta=2$
(unitary) indicates the presence of extended states (Fig.5a)
and the Poisson-like distribution (Fig.5c) shows the existence
of uncorrelated localized states for strong disorder.
The case of edge states is also studied, in which situation the
distribution  can be fitted well with the
Gaussian function (Fig.5b).
Every inset shows a representative distribution of local currents for 
each case.
Very illustrative is Fig.6 which shows the variance
$\delta t_n$ for all level spacing of the Hamiltonian (1).
One may learn that:
a) at $W < 4$,
for most of the states,  $\delta t_{n} \approx 0.42$, 
 which is the typical
value for WD distribution with $\beta=2$. b) for larger $W$, the
variance increases towards $\delta t_{n}= 1.0$ specific to the 
Poisson distribution.
This value cannot practically be reached because of finite dimension of
the plaquette. c) inbetween, at $W\approx 6$, the variance equals 0.52,
and the probability distribution is well-fitted by
the WD function
with $\beta=1$ (so, here
the influence of the magnetic field is lost and the system behaves
'orthogonal'). d) the lowest states, originating from the first band
( n=1,.,6 ) get localized faster than the others,
while the states from the first gap (n=7,.,10) are very
robust against the localization process.
%f6
%\vspace{-4.742cm}
%\vspace*{2cm}
\begin{figure}
\vspace*{-4.2400cm}
\epsfxsize 95mm
\epsffile{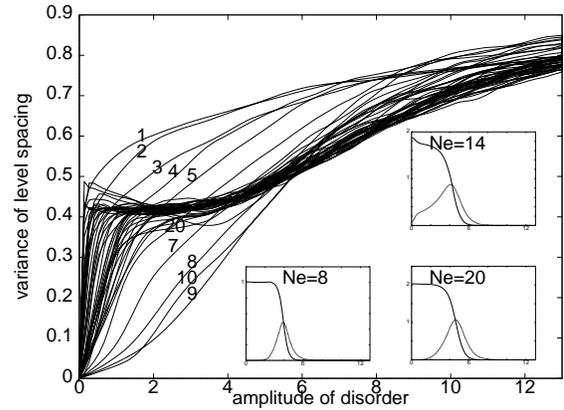}
\vspace*{-2.50cm}
\caption{The variance of level spacing $\delta (t_n)$ (n=1,..,50) 
vs. amplitude of disorder $W$. N=M=10.  
 The insets show the evolution
of  $\sigma_{xx}$ and $\sigma_{xy}$ at a given number of electrons 
$N_e$ vs. $W$ (for the same plaquette  coupled to semiinfinite leads).
}
\label{Fig.6}

\end{figure}

%-------------------transport----------------------------------------
At last we discuss the transport properties; for this purpose 
the Landauer-B\"{u}ttiker formalism and the technics from Ref.9
are used.
The conductance in the three regimes (IQH, metal and insulator) may be 
correlated with the spectrum characteristics of the isolated system
discussed above.
Since the edge states are responsible for IQH effect and they are
robust against disorder, this regime survives also in the presence
of disorder as long as $W$ is not too high.  
In what concerns the transmittances between different leads,
it is well-known that
the only non-vanishing ones are $<T_{\alpha\alpha+1}>$, which connect 
consecutive leads ($\alpha$=lead index) and equal an integer value.

The metallic regime occurs when the transport process takes place on the
states which are extended over the whole plaquette. In this case
also the local
current is distributed on the whole area allowing for a non-zero
transmission probability between any pair of leads. This behaviour 
corroborates with the multi-fractal properties of the local
density of states \cite{JMZ}. The metallic region can be crossed in 
different ways. Let assume a constant disorder (say $W$ =1.0) 
and change the Fermi level ;
in this case,  metallic regions are intercalated between the QH regions,
excepting the lowest one which ensures the transition between QH1 and
the insulating phase.
The consequence is that, 
when crossing the metallic region,
the Hall conductance $\sigma_{xy}$ will have a smooth decay between
consecutive plateaus, while $\sigma_{xx}$ will  differ
from zero and have a maximum in the middle of the transition.
In Fig.4, one has to notice that $R_{xx}$ equals $R_{xy}$ in two places:
1) at the transition between QH1 and insulator, where one has also
$\sigma_{xx}\approx 1/2$ in agreement with Huo at al \cite{5},
and 2) near the centre of the spectrum where the system behaves already
classically.

Another way to look at the metal-insulator transition consists in
crossing the metallic zone by increasing the amplitude of disorder $W$,
while  keeping constant the number of electrons ($N_e$) or the Fermi level.
Now the transition is of the type $\nu\rightarrow 0$ and exhibits
maxima of the longitudinal conductance at a critical disorder $W_{c}$
where, as in Ref.8 , the condition $\sigma_{xx} = \sigma_{xy}$  is fulfilled.
Such kind of transitions are shown in the insets of Fig.6 for $N_e$=8
(i.e. in the first gap) and for $N_e$=20 (i.e. in the second gap). 
One may
ask what is going on when $N_e$  corresponds to a band ; this is the
situation for  the third inset $N_e$=14. It can be observed that $\sigma_{xx}$
is different from zero even at small $W$, meaning that the bulk states are
the first ones which become extended under the influence of disorder.
The dotted line which crosses transversally the spectrum in 
the metallic region of Fig.1
represents the critical disorder $W_c$ corresponding to each $N_e$.

In the insulating phase, when the Poisson distribution of the level spacing is
installed, the Hall and longitudinal conductances tend to
zero and the longitudinal resistance increases exponentially.
The transport of the electron through the plaquette is 
performed by tunneling on localized states.

%.............conclusions..........................
 
In conclusion, the metallic regime  is characterized by a
Wigner-Dyson distribution of the  level spacing with
$\beta=2$ (unitary ensemble). 
As the system evolves towards 
insulator, the orthogonal WD distribution ($\beta=1$) that showes up 
at a given higher disorder indicates the loss of influence of the magnetic 
field.  Simultaneously, the magnetization decays to zero.
The QH phase is characterized by a Gaussian distribution of level spacing.
Due to the different chirality of the edge and extended states, 
when crossing the metallic zone (at relatively small disorder) 
the magnetization
of the ground state shows a toothsaw behavior as function of the 
filling factor similar to experimental results.

Acknowledgments.
A.A. is very grateful to Professor Johannes Zittartz for his
hospitality at Institut f\"{u}r Theoretische Physik der Universit\"{a}t
K\"{o}ln where this work was partially performed under SFB 341.
Illuminating discussions with J.Hajdu, B.Huckestein, M.Zirnbauer,
K.Maschke and A.Manolescu are thankfully acknowledged. We thank
Romanian Academy for the support under the grant No.69/1999.

\noindent$^*$Permanent address: National Institute of Materials Physics
POBox MG7, Bucharest-Magurele, Romania.

\end{document}